\documentclass[11pt]{article}
\usepackage{epsfig,cite}
\usepackage{graphicx}
\usepackage{multicol}
\usepackage{color}
\usepackage{amssymb}

\def\sv{\langle \sigma v\rangle_0}

\def\bea{\begin{eqnarray}}
\def\eea{\end{eqnarray}}

\def\dbar{{\overline{D}}}
\def\sv{{\langle\sigma v\rangle_0}}
\def\units{{{\rm m}^{-2}{\rm s}^{-1}{\rm sr}^{-1} {\rm GeV}^{-1}}}
\def\unitsn{{{\rm m}^{-2}{\rm s}^{-1}{\rm sr}^{-1} {\rm (GeV/n)}^{-1}}}
\newcommand\prd[3]   
		{{Phys.\ Rev.\ }{\bf D #1} (#2) #3}
\newcommand\prl[3]   
		{{Phys.\ Rev.\ Lett.\ }{\bf #1} (#2) #3}
\newcommand\plb[3]   
		{{Phys.\ Lett.\ }{\bf B #1} (#2) #3}
\newcommand\npb[3]    
		{{Nucl.\ Phys.\ }{\bf B #1} (#2) #3}
\newcommand\app[3]   
		{{Astropart.\ Phys.\ }{\bf #1} (#2) #3}
\newcommand\jhep[3]  
		{{J. High Energy Phys.\ }{\bf #1} (#2) #3}
\newcommand\epjc[3]  
		{{Eur.\ Phys.\ J. }{\bf C #1} (#2) #3}
\newcommand\npps[3]  
		{{Nucl.\ Phys.\ }{\bf #1} {\it(Proc.\ Suppl.)} (#2) #3}
\newcommand\jcap[3]  
		{{JCAP\ }{\bf #1} (#2) #3}

\def\sss{\scriptscriptstyle}

\begin{document}
\begin{titlepage}
\pagestyle{empty}
\baselineskip=21pt
\rightline{FSU--HEP--051030}
\vskip 0.9in
\begin{center}
{\huge\sf
\mbox{Low energy antideuterons: shedding light on dark matter}}\\[0.4cm] 
\end{center}
\begin{center}
\vskip 0.5in
{\Large\sf Howard Baer $^{a}$ and Stefano Profumo $^{b,a}$}\\
\vskip 0.2in
{\it {(a) Department of Physics, Florida State University, 
Tallahassee, FL 32306, USA}}\\
{\it {(b) California Institute of Technology, Mail Code 106-38, Pasadena, CA 91125, USA}}\\
{E-mail: {\tt baer@hep.fsu.edu, profumo@caltech.edu}}\\
\vskip 0.4in
{\bf Abstract}
\end{center}
\baselineskip=18pt \noindent

\noindent Low energy antideuterons suffer a very low secondary and tertiary astrophysical background, while they can be abundantly synthesized in dark matter pair annihilations, therefore providing a privileged indirect dark matter detection technique. The recent publication of the first upper limit on the low energy antideuteron flux by the BESS collaboration, a new evaluation of the standard astrophysical background, and remarkable progresses in the development of a dedicated experiment, GAPS, motivate a new and accurate analysis of the antideuteron flux expected in particle dark matter models. To this extent, we consider here supersymmetric, universal extra-dimensions (UED) Kaluza-Klein and warped extra-dimensional dark matter models, and assess both the prospects for antideuteron detection as well as the various related sources of uncertainties. The GAPS experiment, even in a preliminary balloon-borne setup, will explore many supersymmetric configurations, and, eventually, in its final space-borne configuration, will be sensitive to primary antideuterons over the whole cosmologically allowed UED parameter space, providing a search technique which is highly complementary with other direct and indirect dark matter detection experiments.

\vfill
\end{titlepage}

\section{Introduction}

A variety of data now points conclusively to the existence of cold dark matter (CDM) in the universe\cite{Spergel:2003cb}.
The imperative now is to experimentally detect and identify the CDM particle(s)\cite{Jungman:1995df}. 
It is hoped that CDM may be identified in either direct searches, indirect searches, and/or
searches at collider experiments\cite{dicsearches}.
In particular, the possibility of revealing the presence of an exotic particle population in our Galaxy 
through cosmic rays searches has long been envisaged \cite{Zeldovich:1980st}. 
It was soon after realized that if the ``missing matter'' was made up of stable neutralinos, or, more generally, of a pair-annihilating weakly interacting massive particle (WIMP), one could hope to indirectly detect it through gamma rays \cite{gammas}, positrons \cite{Silk:1984zy,Stecker:1985jc,Tylka:1989xj} and through the low-energy production of antiprotons \cite{Silk:1984zy}. 

Though early studies mainly focused on the possibility of tracing anomalies in the observed antimatter spectra back to an exotic contribution from neutralino annihilations in the galactic halo \cite{Tylka:1989xj,Kane:2001fz,Baltz:2001ir}, the possibility of {\em constraining} supersymmetric models through their antimatter yields was also outlined \cite{Baltz:1998xv,Bergstrom:1999qv,Donato:2003xg}. Uncertainties in the background estimations, together with the rather featureless positron and antiproton spectrum predicted in most supersymmetric setups, 
 plague, however, the possibility of ruling out a neutralino component from a given supersymmetric model \cite{Bottino:1998tw,Donato:2003xg}. 
Nevertheless, under the hypothesis that the neutralino component is {\em subdominant} with respect to the background component, 
and assuming that the latter is accurately known, it might indeed be possible to work out whether current data already 
exclude a given supersymmetric model, or if future experiments will be sensitive to the induced antimatter fluxes\cite{Profumo:2004ty,Profumo:2004ff,Hooper:2004bq}.

A critical issue in the discrimination of an exotic component in the cosmic ray spectra resides evidently in the evaluation of the 
abundances of the species under consideration generated by ``standard'' astrophysical processes. 
The computation of the flux of secondary antiprotons produced in high-energy collisions of cosmic ray nuclei with the interstellar gas, 
for instance, is a complex task, and, despite the wealth of experimental information collected in the last decade \cite{bess1,bess2,capricepbar,mass91,heat,capriceeplus}, 
the estimates from various groups differ significantly (see e.g. Fig.~3 in Ref.~\cite{bess1}). 
The production, annihilation and scattering cross sections, the model of antiproton propagation in the Galaxy, 
and the effects of the heliosphere modulation are major sources of uncertainty \cite{Moskalenko:2001ya,Donato:2001sr}. 
In any case, a general consensus has been reached on the fact that the low-energy tail of the secondary antiproton 
spectrum is abundantly populated \cite{Bergstrom:1999qv}, contrary to earlier estimates \cite{Silk:1984zy}. 
Despite the fact that antiprotons cannot be produced at rest, for kinematical reasons, 
various processes contribute to replenishing the antiproton population featuring a low kinetic energy 
(say, below the maximal differential flux, at 2 GeV), including ionization losses, 
a tertiary antiproton population, and inelastic but non-annihilating scattering off the hydrogen atoms in the galactic disk. 
On top of that, the net effect of solar modulation is to further shift the energy spectrum towards lower energies. 
As a result, low energy antiprotons, which are typically abundantly produced in WIMP pair annihilations\cite{Bergstrom:1999qv,Bottino:1998tw}, 
cannot be regarded as a viable window on where to look for a clean new physics signature.

In Ref.~\cite{Donato:1999gy} it was first pointed out that the picture could be totally different in the case of antideuterons, 
the nuclei of the anti-deuterium. 
Donato et al.~\cite{Donato:1999gy} showed that the flux of antideuterons produced in neutralino pair annihilations, 
despite being several orders of magnitude smaller than the antiproton flux, could be much larger than the secondary background, 
estimated in an earlier paper by Chardonnet et al., Ref.~\cite{Chardonnet:1997dv}, 
in the low kinetic energy tail of the spectrum (typically, at kinetic energies per nucleon less than 1 GeV). 
The main reason for the qualitative difference from the case of antiprotons is that the kinetic energy threshold 
for the production of an antideuteron from the collision of an energetic proton on the interstellar medium (ISM) 
is much larger than that needed to create an antiproton. 
Further, energy loss mechanisms are less efficient in shifting the antideuteron energy spectrum towards low energies: 
in particular, given the low antideuteron nuclear binding energy $B\sim2.2$ MeV, the inelastic but non-annihilating 
reactions which efficiently dump the energetic antiprotons were believed to be ineffective in the case of antideuterons, 
which were supposed to be mostly fragmented in those processes.

The detection of even a single antideuteron, provided the secondary background is indeed suppressed, 
looked to be therefore a very promising possibility to have clean evidence for a connection between new physics and cosmic rays. 
Following this result, a clever experimental setup was designed and proposed in Ref.~\cite{Mori:2001dv}, 
consisting of a ``{\em Gaseous Antiparticle Spectrometer}'' (GAPS), designed to capture low energy antinuclei, 
suitably slow them, and detect the subsequent X-ray de-excitation cascade of the exotic atoms where a shell electron is replaced by the antimatter nucleus, 
and finally the nuclear disintegration. 
The peculiar features of all of these processes for each antinucleus species allow in principle to achieve an 
incredibly high discrimination against protons, to the level of 1 part in $10^{12}$ \cite{Mori:2001dv}. 
It was then shown in several papers that antideuteron detection, especially through the GAPS apparatus, 
could provide an extremely efficient indirect detection technique for neutralino dark 
matter\cite{Provenza:2005nq,Baer:2005ky,Baer:2005zc,Baer:2005bu,Baer:2004fu,Masiero:2004ft,Baltz:2002ei,Edsjo:2004pf}. 

A certain number of significant developments have recently occurred regarding antideuteron searches, 
which motivate the present critical re-assessment of the experimental and theoretical prospects for low-energy antideuterons as a dark matter indirect detection technique. 
First, a few months ago the first limit on the cosmic ray antideuteron abundance was set by the BESS collaboration\cite{Fuke:2005it}. 
Remarkably enough, the mentioned upper limit constrains low energy antideuterons (see Sec.~\ref{sec:exp}), 
and provides bounds on the cosmological primordial black holes abundance \cite{Barrau:2002mc,Fuke:2005it}, 
though it is, apparently, too loose to give any constraint on WIMP models \cite{Fuke:2005it}. 
Second, the GAPS experiment has recently undergone a significant R\&D phase, and long duration balloon-borne prototypes 
should be ready within a few years \cite{Hailey:2005yx,hailey,hailey_privcom,Hailey:2003xh}. 
Depending on the experimental setup and on the type of mission, these preliminary launches can have significant scientific opportunities, 
which is certainly worthwhile to assess. On the theoretical side, a new evaluation of the antideuteron background has been carried out 
in \cite{Duperray:2005si}. Previously neglected antideuteron production and energy loss processes, 
including secondary antideuteron production from antiproton scattering off the ISM, and a tertiary antideuteron component, 
originating from the previously neglected non-annihilating inelastic scattering processes, 
have been shown to largely populate the low-energy end of the antideuteron spectrum. 
Although large uncertainties, related to both the nuclear reactions and the propagation and solar modulation effects
somewhat blur the final result, the main message is that the ``background'' at low energies might not be as low as previously thought, 
and should be taken into account when assessing the possibility of gaining a clean indication of an exotic component. 
Lastly, the existing analyses of antideuterons produced by WIMP annihilation only refer to neutralino dark matter, 
while other dark matter candidates have been recently proposed, and investigated with respect to their potential direct and indirect signatures. 
An incomplete list includes the lightest Kaluza-Klein particle (LKP) of universal extra-dimensional scenarios\cite{Appelquist:2000nn,Servant:2002aq}, 
and the right-handed Kaluza-Klein neutrino of warped 5-dimensional grand unified theories (GUT) 
with a conserved $Z_3$ parity \cite{Agashe:2004ci,Agashe:2004bm,Hooper:2005fj} (LZP).

The goals of the present note are therefore to {\it i}.) sketch the current and projected experimental status for antideuteron searches (Sec.~\ref{sec:exp}), 
{\it ii}.) to present calculational details for both the signal and the recently re-evaluated low energy background (Sec.~\ref{sec:calcdet}- \ref{sec:bckg}), 
{\it iii}.) to evaluate the flux of antideuterons in a wide range of WIMP dark matter scenarios, while 
assessing the prospects of detection at the various upcoming experiments, including the issue of the background (Sec.~\ref{sec:wimpmodels}) 
and to compare these results against prospects for other direct and indirect dark matter detection experiments (Sec.~\ref{sec:compare}), 
and, finally, {\it iv}.) to give a realistic picture of the uncertainties 
involved in the antideuteron flux computation (Sec.~\ref{sec:unc}).

\section{Antideuteron searches: experimental status and prospects}\label{sec:exp}

Antideuteron searches can be performed either with magnetic spectrometers mounted on balloon-borne (BESS/BESS-Polar) or space-borne (AMS) missions, 
or through GAPS-like devices, based on the radiative emissions of antiparticles captured into exotic atoms. 
The latter can be installed again either on balloons or on satellites, and are specifically designed to look for low-energy antiparticles.

The BESS experiment looked for low-energy antideuterons during four flights (1997, 1998, 1999, and 2000), 
in the kinetic energy interval $0.17\div 1.15$ GeV/n. The upper and lower kinetic energy limits come, respectively, 
from the particle identification procedure and from the decrease of geometrical acceptance and mean free path through the detector. 
Without assumptions on the $\dbar$ spectrum shape, the BESS collaboration, by combining all four missions, 
derived an upper limit on the $\dbar$ flux at 95\% C.L.,  of
\begin{equation}\label{eq:besslimit}
\phi^{\rm BESS}_{\dbar}\ < 1.9\times 10^{-4}\ \unitsn.
\end{equation}
The Fisk solar modulation parameter $\Phi$ was derived from the $p$ data from the same experiment, 
and set to 500, 610, 648, 1334 MV. In \cite{Fuke:2005it}, it was claimed, based on the results of Ref.~\cite{Donato:1999gy}, 
that no constraints from this result apply to the case of  $\dbar$s produced in WIMP annihilations, 
while the limit to the flux places a stronger bound on the abundance of primordial Black Holes\cite{Barrau:2002mc}. 
The latter is, in any case, weaker by two orders of magnitude than what was obtained from $\overline p$ data\cite{Fuke:2005it}. 

The computation of the sensitivity of the AMS-02 payload\cite{ams} to low-energy antideuterons 
involves the problem of a careful treatment of the geomagnetic cutoff effects on the impinging particles' flux at the 
particular international space station orbit. This issue has been addressed in \cite{Donato:1999gy}, 
where, for a total data-taking time of 3 yrs ($\sim 10^8$ s), and for an antideuteron kinetic energy band extending 
from the AMS threshold of 100 MeV/n to $\sim2.7$ GeV/n, the inferred acceptance reads $5.5\times 10^7\ {\rm m}^2\ {\rm s}\ {\rm sr}\ {\rm GeV}$. 
This can be translated into a critical flux of antideuterons at a given $\dbar$ kinetic energy\cite{Donato:1999gy}, 
but, in a model independent approach, we shall compute the actual number of expected primary antideuterons $N_{\dbar}$ 
detected in the AMS energy range, using the above quoted acceptance, and declare a model above or below the AMS sensitivity if the resulting $N_{\dbar}\gtrless1$.

As mentioned in the introduction, the GAPS experiment has recently undergone a rich phase of R\&D, carried out at the KEK accelerator in Japan. 
Various target materials were analyzed, including solid and liquid targets (which led the collaboration to change the experiment's 
acronym to ``General'' instead of ``Gaseous'' A.P.S.). The results of these preliminary tests have been reported in various 
conferences\cite{Hailey:2005yx,hailey,hailey_privcom,Hailey:2003xh}, and look very promising. 
In particular, it has been realized that solid and liquid targets can greatly simplify the needed payload mass (thanks to the removal of the dead mass of the gas handling system) and the complexity of the apparatus, 
yielding an increased background rejection capability enabling the capture of more than 3 X-rays, as initially conceived. 
Further, pion showers ($\pi^*$) and nuclear X-rays from the antiparticle annihilation in the target nuclei, neglected in the original sensitivity calculations\cite{Mori:2001dv}, 
have been shown to significantly increase the antiparticle identification capability. 
The GAPS collaboration then plans to test the finalized payload with a prototype as soon as 2009, and to achieve a long duration balloon (LDB) 
flight from Antarctica, or an ultra-LDB (ULDB) flight from Australia as soon as 2011 \cite{Hailey:2005yx,hailey,hailey_privcom}. 
A preliminary evaluation of the sensitivity of the two balloon-borne options (with the LDB sensitivity based on 3 flights) gives \cite{hailey_privcom}
\begin{equation}
\phi^{\rm LDB}_{\dbar}\ \simeq 1.5\times 10^{-7}\quad \units \ {\rm and}\quad \phi^{\rm ULDB}_{\dbar}\ \simeq 3.0\times 10^{-8}\ \units
\end{equation}
over a bandwidth of $0.1<T_{\dbar}/({\rm GeV/n})<0.25$ \cite{hailey_privcom}. 
The sensitivity of the GAPS prototype designed in \cite{Mori:2001dv} mounted on a satellite was instead assessed in \cite{Mori:2001dv}, 
keeping track of the geomagnetic effects mentioned above. The recent accelerator testing of the GAPS prototype \cite{Hailey:2005yx} essentially confirm the sensitivity quoted in \cite{Mori:2001dv}. For a 3 yr mission, the projected sensitivity reads
\begin{equation}
\phi^{\rm GAPS/S}_{\dbar}\ \simeq 2.6\times 10^{-9}\quad \units,\quad\quad {\rm for} \quad\quad  0.1<T_{\dbar}/({\rm GeV/n})<0.4 .
\end{equation}
A last, very optimistic option, mentioned in \cite{Mori:2001dv}, is to send GAPS on a probe in deep space, where eventually solar modulation 
effects can be significantly reduced. Depending on the spectral shape of the differential $\dbar$ yield from WIMP pair annihilation, solar modulation can deplete the low energy antideuteron flux, hence this interplanetary GAPS setup might represent the ultimate probe for DM searches via detection of low energy $\dbar$s \cite{Profumo:2004ty}.

\section{Searching for WIMP-annihilation induced antideuterons}\label{sec:wimps}
\subsection{Primary (WIMP-induced) antideuterons: calculational details}\label{sec:calcdet}

\begin{figure}[!t]
\begin{center}
\epsfig{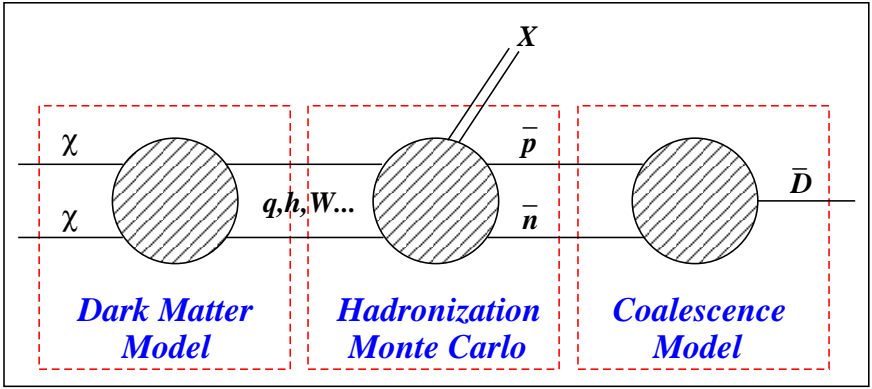}
\end{center}
\caption{\it\small 
A conceptual outline of how to compute the antideuteron flux induced by a dark matter particle ($\chi$) pair annihilation. 
First, the dark matter particle physics model provides the pair annihilation cross section into standard model particles ({\em e.g.} 
a quark-antiquark pair, gauge and/or Higgs bosons etc.). Then, a Monte Carlo hadronization simulation translates the elementary particle 
output into the flux of antiprotons and antineutrons. 
Finally, a nuclear physics model (in the present case the {\em coalescence} model) provides the final yield of antideuterons.}
\label{fig:graphic}
\end{figure}
The computation of the differential flux of $\dbar$s per kinetic energy per nucleon interval induced by WIMP pair-annihilations involves a number of steps, 
which we briefly review below, referring the reader to \cite{Donato:1999gy,Profumo:2004ty} for more details. 
To frame the discussion, we give a conceptual sketch of the ingredients involved in the assessment of the dark matter induced antideuteron flux in Fig.~\ref{fig:graphic}. 
We also outline below a {\em reference setup}, that will be used in Sec.~\ref{sec:unc} to evaluate the various sources of uncertainty 
which plague the computation outlined below.

The computation of the source spectrum for the primary antideuteron flux originating from WIMP pair annihilation is based on three hypothesis: 
(1) the probability of producing a pair of antinucleons is given by the product of the probability of producing a single antinucleon ({\em factorization}) 
(2) the antineutron production cross section is equal to the antiproton production cross section ({\em isospin invariance}) and 
(3) the formation of an antideuteron can be described by the {\em coalescence model}. 
We refer the reader to Ref.\cite{Chardonnet:1997dv,Donato:1999gy,Donato:2001sr,Duperray:2005si} for a through discussion of the validity of these assumptions. 
In particular, we stress that the factorization assumption is conservative, in that the probability of pair producing antinucleons in the same jet 
is presumably not factorized, since their momenta will not be isotropically distributed. 
The main idea of hypothesis (3) is that whenever the difference of the momenta of an antiproton and an antineutron produced in a jet 
resulting from a WIMP pair annihilation is less than a phenomenologically given value $2p_0$, where $p_0$ indicates the {\em coalescence momentum}, 
then an antideuteron is formed. The differential energy spectrum of primary antideuterons produced in the pair annihilation of a WIMP $\chi$ 
can then be expressed by \cite{Donato:1999gy}
\begin{equation}
\frac{{\rm d}N_{\dbar}}{{\rm d}E_{\dbar}}=\left(\frac{4\ p_0^3}{3\ k_{\dbar}}\right)\ \left(\frac{m_{\dbar}}{m_{\bar p}\ m_{\bar n}}\right)\ 
\sum_{f}BR(\chi\chi\rightarrow f)\times\left(\frac{{\rm d}N^{(f)}_{\bar p}}{{\rm d}E_{\bar p}}\left(E_{\bar p}=E_{\dbar}/2\right)\right)^2 ,
\end{equation}
where $E_{\dbar}^2=m_{\dbar}^2+k_{\dbar}^2$, $f$ indicates any final state of the WIMP pair annihilation process occurring with a branching ratio 
$BR(\chi\chi\rightarrow f)$, and ${\rm d}N^{(f)}_{\bar p}/{\rm d}E_{\bar p}$ is the antiproton differential yield for the final state $f$. 
The latter is computed using the results of the {\tt Pythia} Monte Carlo event generator \cite{pythia}, as implemented in the {\tt DarkSUSY} package\cite{Gondolo:2004sc}. 
The source spectrum is then specified at every point in the galactic halo once the shape of the DM halo itself is given. 
We take here as a reference model the adiabatic contraction \cite{blumental} of the N03 halo profile \cite{n03} (see Ref.~\cite{pierohalos} for details), 
which closely resembles the profile proposed by Moore et al., \cite{Moore:1999nt}. The particular configuration for the dark matter halo we use here has been obtained after implementing all available dynamical constraints and numerical simulations indications on the halo-mass concentration correlation \cite{pierohalos}. The local halo density at the Sun location for the dark matter halo under consideration reads $\rho^{\rm loc}_{\rm DM}\simeq0.38\ {\rm GeV/cm}^3$ (see Sec.~\ref{sec:unc} for a discussion of the other dark matter halos we considered and the corresponding local halo densities). We consider a smooth halo profile, but we discuss in Sec.~\ref{sec:unc} the effects of DM halo substructures. 

The reference value we assume for the coalescence momentum is 58 MeV, the same choice as in Ref.~\cite{Chardonnet:1997dv} and \cite{Donato:1999gy}, 
not too far from what is expected from the antideuteron binding energy, $\sqrt{m_p\ B}\approx 46$ MeV. 
Again, see Sec.~\ref{sec:unc} for a discussion of the uncertainties on the primary flux generated by a range of viable coalescence momenta.

We sketch the effects of propagation of antideuterons through the Galactic magnetic fields in a two-dimensional diffusion model in the steady state 
approximation (see Ref.~\cite{Bergstrom:1999qv} for details). Following \cite{Moskalenko:2001ya}, the diffusion region is taken to be cylindrical, 
with a radius of 30 kpc and half-height $h_h=4$ kpc, plus a galactic wind term with velocity $v_w=10$ km/s. 
Reacceleration effects are mimicked through a diffusion coefficient $D$ which contains a power law behavior as a function of rigidity $R$, 
and a constant value below the critical rigidity $R_0$, namely
\begin{equation}
\nonumber D=D_0(R/R_0)^{0.6}\qquad R\geq R_0
\end{equation}
\begin{equation}
\nonumber D=D_0\quad\qquad R\leq R_0 ,
\end{equation}
where the various coefficients are again taken from the analysis of Ref.~\cite{Moskalenko:2001ya} to be $D_0=2.5\times10^{28}$ ${\rm cm}^2\ {\rm s}^{-1}$ and $R_0=4$ GV. 
This setup is then interfaced with the semi-analytical diffusive-convective model of Ref.~\cite{Bergstrom:1999qv}, as implemented in {\tt DarkSUSY} \cite{Gondolo:2004sc}.

The solar modulation effects have been accounted for in the framework of the Gleeson-Axford analytical force-field approximation\cite{GleesonAxford}, 
where the interstellar flux at the heliospheric boundary, ${\rm d}\Phi_b/{\rm d}T_b$, and at the Earth, ${\rm d}\Phi_\oplus/{\rm d}T_\oplus$ are related by
\begin{equation}
\frac{{\rm d}\Phi_\oplus}{{\rm d}T_\oplus}(T_\oplus)=\frac{p^2_\oplus}{p^2_b}\ \frac{{\rm d}\Phi_b}{{\rm d}T_b}(T_b),
\end{equation}
where the energy at the heliospheric boundary is given by $E_b=E_\oplus+|Ze|\phi_F$, and $p_b$ and $p_\oplus$ stand for the momenta at the boundary and at the Earth, 
and $\phi_F$ is the solar modulation, or Fisk\cite{fisk} parameter, which is supposed to be, for simplicity, charge-sign independent. 
We take as a reference value $\phi_F=800$ MV, and discuss the effects of the solar activity on the $\dbar$ flux in Sec.~\ref{sec:unc}.

\subsection{Secondary and tertiary antideuterons: the role of the background}\label{sec:bckg}

Galactic antideuterons are secondary products of various reactions involving an energetic particle inelastically scattering off a target nucleon or nucleus. 
The $\dbar$ formation is then described in the framework of the above described coalescence model. 
The antideuteron flux has been recently re-evaluated in Ref.~\cite{Duperray:2005si}. 
It has been shown that for kinetic energies per nucleon larger than around 1 GeV, the dominant antideuteron production processes are 
$(p\ p)$, $(p\ He)$ and $(He\ p)$ reactions, with the first particle acting as projectile and the second as target. 
The dominant processes between 0.3 and 1 GeV are instead found to be $(\overline p\ p)$ and $(\overline p \ He)$ reactions, previously neglected. 
The energy loss induced by the elastic scattering at energies below 0.5 GeV involves small momentum transfers, and it is found to be negligible. 
However, it was also noticed in Ref.~\cite{Duperray:2005si} that the non-annihilating inelastic process 
\begin{equation}
\overline D\ p\ \rightarrow\ \overline D\ X
\end{equation}
may involve large energy momentum losses of the scattered particles, and induces a ``{\em tertiary}'' component in the region where the 
secondary flux is extremely suppressed. The tertiary component gives a flat source contribution to the interstellar $\dbar$ flux, below 0.5 GeV, of approximately
\begin{equation}\label{eq:dbarbckg}
\phi_{\dbar}^{\rm ter}\simeq 3-6\times 10^{-9} \unitsn,
\end{equation}
depending on the assumed coalescence model\cite{Duperray:2005si}. 
The effect of solar modulation is then to re-shape the $\dbar$ flux spectrum to a power law behavior in the low energy regime. 

A further source of antideuterons is the atmospheric production when a cosmic ray interacts with the Earth atmosphere. 
For balloon-borne experiments, this component must be taken into account. 
For satellite-borne experiments, the atmospheric background can still manifest itself through the bending of the trajectories 
of particles created in the atmosphere, but it can be separated from the galactic flux on dynamic and kinematic grounds\cite{stormer}.

Turning to the low energy range, we compute here the galactic $\dbar$ background for AMS and for GAPS, 
integrating the fluxes computed in Ref.~\cite{Duperray:2005si}. 
Neglecting the tertiary component and the secondary component induced by energetic $\bar p$ , 
the background in the low energy range is negligible in both cases \cite{Donato:1999gy}. 
Taking those components into account, instead, translates in a number of background events (at $\phi_F=500$ MV and 1000 MV, respectively) 
of 0.04 (0.05) events for the ULDB GAPS experiment, of 1.4 (1.8) events for the GAPS setup on a satellite and of 2.1 (2.4) events for AMS. A GAPS detector on an interplanetary probe would suffer a background of 0.77 events (independently of the solar modulation parameter). The atmospheric background, as simulated in \cite{Duperray:2005si}, is suppressed with respect to the galactic background by at least one order of magnitude. The uncertainties related to the background computation are certainly very large, 
and a further assessment of the range of possible variations of the astrophysical secondary and tertiary $\dbar$ component would by all means be desirable for forthcoming dedicated search experiments. We take here as our {\em reference background model} the background at $\phi_F=800$ MV (we discuss uncertainties on the background and primary $\dbar$ flux in Sec.~\ref{sec:unc}).

Taken at face value, the above quoted background levels give a picture of the prospects for using low-energy antideuteron detection as a 
clean signature of an ``exotic'' cosmic species which is significantly different from the original one given in Ref.~\cite{Donato:1999gy}, where the detection of even a single low-energy antideuteron in any search experiment could be regarded as a ``new physics'' signature. 
More quantitatively, the probability distribution for the detection of $N$ background antideuterons will follow a Poissonian distribution, 
with mean set to the background levels $b$ quoted above, $P(N,b)=(b^N e^{-b})/N!$. 
This implies that the probability of detecting one or more background events at the ULDB GAPS mission is less than 5\%. 
On the other hand, the probability of detecting more than 5 events at GAPS/satellite, 
and 6 events at AMS is less than 4\%. There is an 82\% chance of detecting 1 to 5 background events at GAPS/satellite, and a 90\% chance of 
detecting 1 to 6 events at AMS. 
Clearly, the detection of a single antideuteron at the GAPS/balloon experiment would indeed give a strong indication of a new physics signal, 
while the detection of even 4 or 5 events, either at AMS or at GAPS/satellite would only marginally demonstrate that an exotic source is in place. 

The possibility of deriving a {\em lower limit} on the source signal $s>0$ at, say, 95\% C.L., under the hypothesis of a reliable computation of the background level $b$, may proceed solving the equation
\begin{equation}
\sum_{n=0}^{N-1}\ P(n,x)=0.95 .
\end{equation}
The value of $N$ such that $x>b$ will give a lower limit on $s>x-b$ at the 95\% C.L.. This procedure gives then a rule to find the minimal number of low-energy detection events needed to claim, at 95\% C.L., the occurrence of an exotic $\dbar$ source. This approach gives $N=1$ for ULDB GAPS/balloon, while $N=5$ for GAPS/satellite and $N=6$ for AMS. A GAPS detector mounted on an interplanetary probe, for which we use a median value of the interstellar antideuteron background quoted in Eq.~(\ref{eq:dbarbckg}), would instead require $N=3$. In what follows, we will quote the WIMP pair annihilation cross sections needed to produce {\em i.}) at least one primary $\dbar$ and {\em ii.}) a flux of primary antideuterons such that the sum of the primary and of the background component equals the values of $N$ quoted above, for each experimental setup. This corresponds to a flux which should guarantee the statistical discrimination of a non-vanishing primary $\dbar$ component: for instance, in the case of AMS, one has $N=6$ and $b=2.3$ events, hence the primary flux required will be of 3.7 events, and analogously for the case of GAPS.

\subsection{$\dbar$ flux from WIMP annihilation in the galactic halo}\label{sec:wimpmodels}
\begin{figure}[!t]
\begin{center}
\epsfig{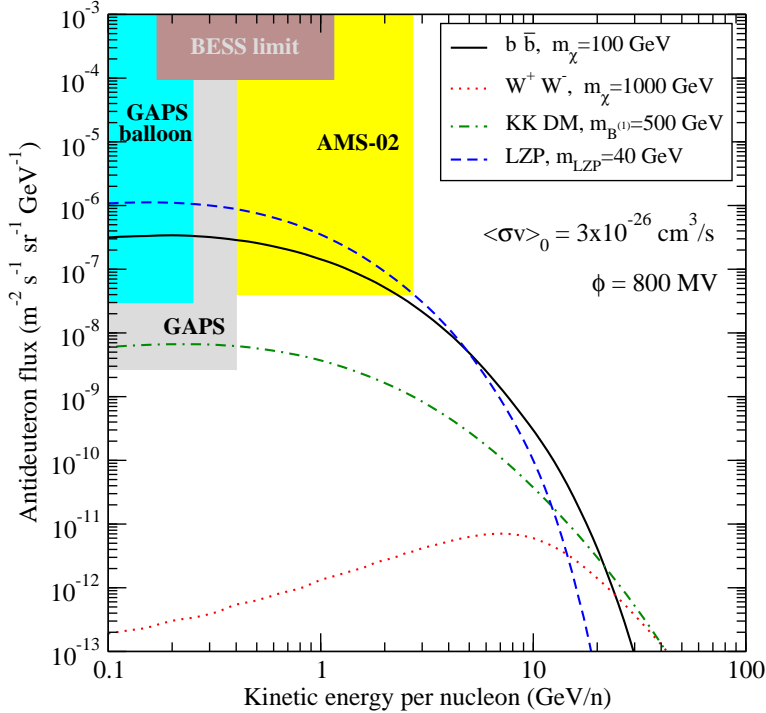}
\end{center}
\caption{\it\small 
Differential antideuteron flux from four different WIMP models, as a function of the antideuterons' kinetic energy per nucleon. 
The solid black line corresponds to a WIMP with mass 100 GeV annihilating with BR=1 into a $b\bar b$ pair, 
the red dotted line to a 1000 GeV WIMP annihilating with BR=1 into $W^+W^-$ pairs, the green dot-dashed line to a 500 GeV $B^{(1)}$ 
(the Kaluza-Klein first excitation of the hypercharge gauge boson), LKP in the UED scenario, while the blue dashed line to a LZP particle 
pair annihilating dominantly through the $Z$ $s$-channel resonance, with a mass of 40 GeV. 
The shaded regions correspond to the sensitivities of various existing and proposed experiments featuring antideuteron searches.}
\label{fig:dnde}
\end{figure}
We collect in Fig.~\ref{fig:dnde} the results of the computation of the antideuteron flux for several models containing WIMP candidates. Throughout this section we implement the propagation and dark matter halo models described in the previous Sec.~\ref{sec:calcdet}, and refer the reader to Sec.~\ref{sec:unc} for a discussion on how our results would be affected by changing those assumptions.
We fix in Fig.~\ref{fig:dnde} the value of the thermally averaged WIMP pair annihilation cross section to $\sv=3\times 10^{-26}\ {\rm cm}^3/{\rm s}$, 
which is indicative (at least for non-coannihilating or resonantly annihilating WIMPs featuring an unsuppressed $s$-wave annihilation cross section) 
of the typical cross section range for models giving a WMAP relic abundance \cite{Spergel:2003cb}, according to the qualitative relation \cite{Jungman:1995df}
\begin{equation}\label{eq:naivesv}
\sv\approx\frac{3\times 10^{-27}\ {\rm cm^3}\ {\rm s}^{-1}}{\Omega_\chi h^2}.
\end{equation}
The first two models feature a single final state, respectively $b\bar b$ and $W^+W^-$, and two different masses, respectively $m_\chi=100$ GeV and $m_\chi=1000$ GeV. 
The choice of the masses is rather arbitrary, but it has been repeatedly shown that, for instance, supersymmetric models with a neutralino LSP mainly 
annihilating into gauge bosons pairs (such as wino- or higgsino-like neutralinos) typically feature a mass in the TeV range. 
On the other hand, the $b\bar b$ final state is often found to be the dominant annihilation channel\cite{bo} for low mass neutralinos (especially at large $\tan\beta$), 
for instance in the minimal supergravity model\cite{msugra}. 
The third model we consider is the $B^{(1)}$ LKP of UED models \cite{Appelquist:2000nn,Servant:2002aq}. 
The branching ratios for this model have been computed in Ref.~\cite{Servant:2002aq}, and the dominant final state channels responsible for antiproton
(and thus $\dbar$s) production are up-type quarks (see also the recent analysis of the anitproton yields for this model in Ref.~\cite{Barrau:2005au}). 
We picked a representative mass of 500 GeV, which should fall in the WMAP preferred mass range, according to recent evaluations 
which also take into account the full impact of coannihilations \cite{Kong:2005hn,Burnell:2005hm}. 
Finally, we consider the LZP particle of the 5-dimensional warped GUT model of Ref.~\cite{Agashe:2004ci,Agashe:2004bm,Hooper:2005fj}, 
in the mass range where it predominantly annihilates through the $Z$ boson $s$-channel resonance. 
The final state products thus follow the $Z$ decay branching fraction (the $Z$ is very close to being on-shell). 
This model can clearly be regarded as a benchmark for any WIMP model where the main annihilation channel is through the $Z$. 
The mass of the LZP was set to 40 GeV. In the Figure, we also shade the sensitivities of current and future antideuteron search experiments, 
as discussed in the preceding Sec.~\ref{sec:exp}.

A few general remarks can be drawn from Fig.~\ref{fig:dnde}. 
First, if the dominant final state channel is into a quark-antiquark pair, the maximal $\dbar$ flux is predicted to occur 
exactly in the low energy range of interest for $\dbar$ DM searches, and it is characterized by a plateau in the range which will be 
explored by a GAPS-like apparatus ($T_{\dbar}\lesssim 0.4$ GeV/n). We also notice that the $\dbar$ spectrum for up-type quarks is slightly harder 
than that for down-type quark, but this difference occurs only at relatively large kinetic energies. 
If, instead, the dominant final state is into a pair of gauge bosons (the case of the pair annihilation into $ZZ$ gives a very similar differential energy spectrum, 
with a factor 2 larger yield in the low energy region) the $\dbar$ yield spectrum is qualitatively different. 
The maximal antideuteron flux is reached at larger energies, and the low energy band is less populated. Further, the overall flux is lower than in the quark-antiquark final state. These effects are traced back to the fact the hadronization products of gauge bosons decays tend to be very energetic, suppressing the overall antideuteron flux (due to the coalescence momentum condition) and yielding, on average, more energetic antideuterons.
At lower WIMP masses, the antideuteron yield from gauge boson pairs at low kinetic energies turns out to be, on the other hand, less suppressed  (see {\em e.g.} Fig.~4 of 
Ref.~\cite{Profumo:2004ty}).
We thus expect that, in general, WIMP models featuring a dominant pair annihilation rate into gauge bosons pairs will be disfavored in the context of $\dbar$ DM searches.

\begin{figure}[!t]
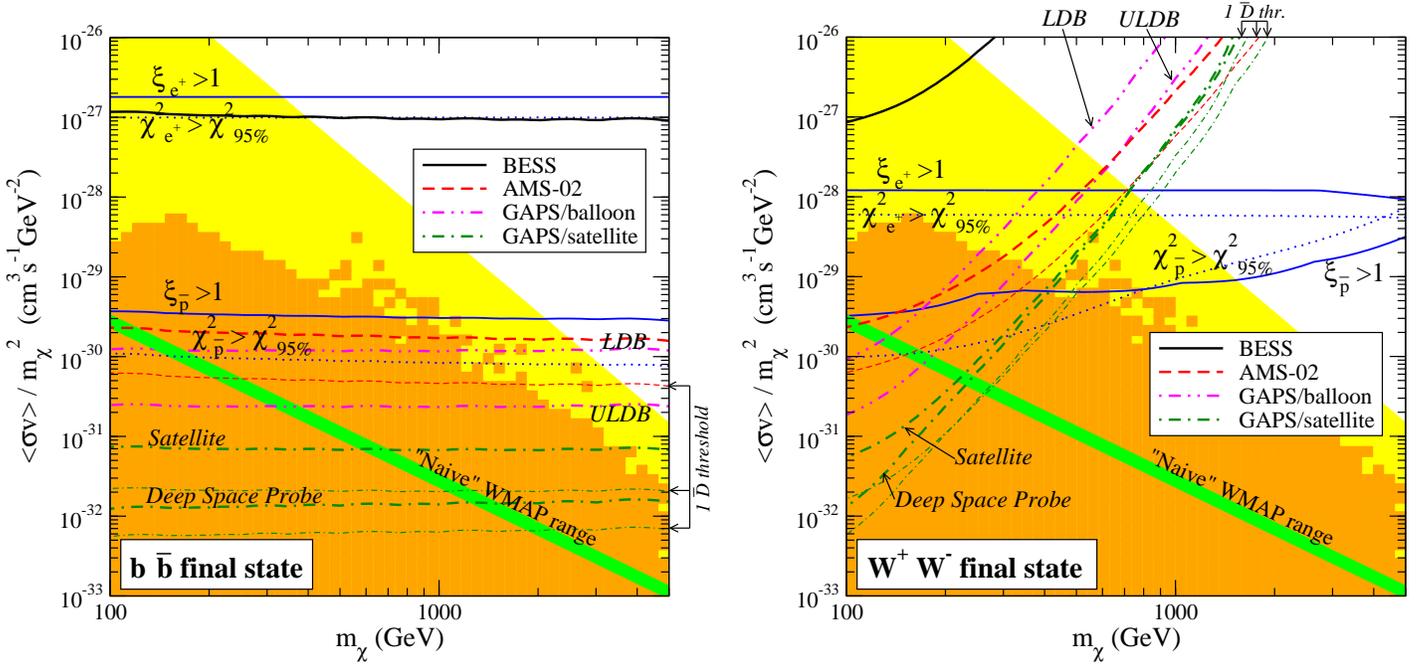

\begin{center}
\hspace*{-1.0cm}\mbox{\epsfig{file=plots/bbbar_new.eps,height=8.5cm}\quad\epsfig{file=plots/ww_new.eps,height=8.8cm}}
\end{center}
\caption{\it\small 
The sensitivity reach of antideuteron search experiments for WIMP pairs annihilating with BR=1 into $b\bar b$ pairs (left panel) and $W^+W^-$ pairs (right panel), 
in the plane defined by the particle mass $m_\chi$ and by the factor $\sv/m_\chi^2$. 
The black solid line gives the experimental upper limit on the antideuteron flux from BESS\cite{Fuke:2005it}. 
For future experiments, the lines correspond to the critical primary antideuteron flux giving an expected number of detected antideuterons 
over the full lifetime of the experiment equal to 1. 
The red dashed line sketches the projected sensitivity of the AMS-02 experiment after 3 years of data taking. The fainter line corresponds to the detection threshold of 1 primary $\dbar$, while the thicker line to a number of primary $\dbar$s sufficient to disentangle them from the background.
The magenta double-dotted-dashed lines indicate the sensitivity of a balloon-borne GAPS setup in a LDB mission over Antarctica (upper line) 
and in a ULDB mission over Australia (lower line). 
The sensitivity of a satellite-borne and interplanetary probe version of GAPS are instead indicated by a dot-dashed green line. Again, the fainter lines correspond to the detection threshold of 1 primary $\dbar$, while the thicker ones to a number of primary $\dbar$s sufficient to disentangle them from the background.
The upper and lower blue lines correspond to the bounds from the measured flux of positrons and antiprotons, respectively. 
Points on the plot above the dotted lines give a total $\chi^2$ for the background plus primary component which is excluded at 95\% C.L., 
while for points above the solid lines, the positron/antiproton fluxes induced by WIMP annihilations only (primaries) exceed, at the 2-$\sigma$ level, 
the experimentally measured flux in at least one energy bin. 
Finally, the orange-shaded region corresponds to supersymmetric models giving a WMAP relic abundance within 2-$\sigma$, 
while the yellow shaded area corresponds to the maximal region spanned by supersymmetric models in the $(m_\chi,\sv/m_\chi^2)$ plane 
(see Ref.~\protect{\cite{Profumo:2005xd}}). As a guideline, 
we also indicate the ``naive'' cross section range favored by the WMAP deduced CDM abundance according to the relation given in Eq.~(\ref{eq:naivesv}).} 
\label{fig:susy}
\end{figure}

\begin{figure}[!t]
\begin{center}
\hspace*{-1.0cm}\mbox{\epsfig{file=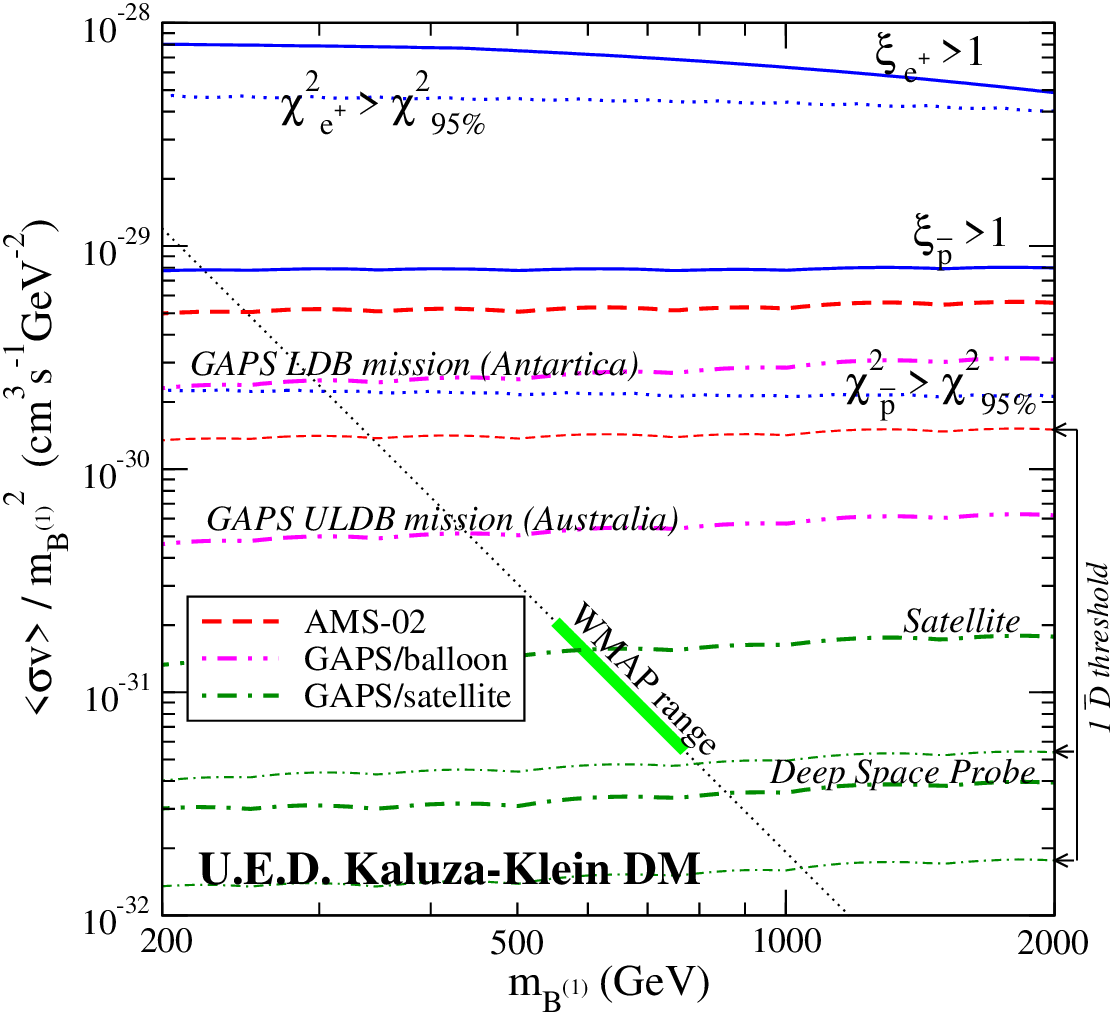,height=8.5cm}\quad\epsfig{file=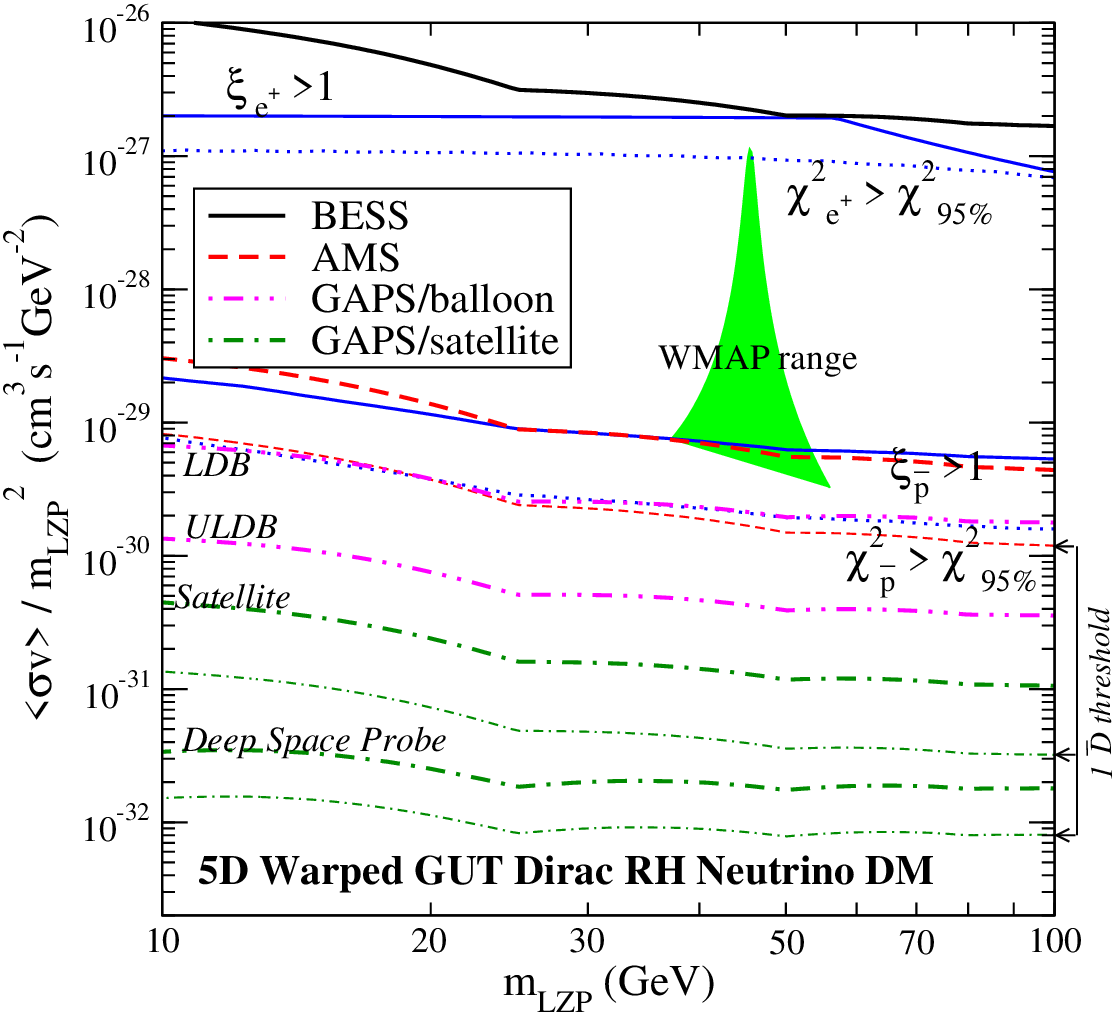,height=8.5cm}}
\end{center}
\caption{\it\small 
The sensitivity reach of antideuteron search experiments for two extra-dimensional DM models: 
the UED model, with a $B^{(1)}$ LKP (left panel) and the warped extra-dimensional GUT scenario of Ref.~\cite{Agashe:2004ci,Agashe:2004bm,Hooper:2005fj}, 
featuring a right-handed neutrino as the LZP, in the mass range where the LZP resonantly annihilates into the $Z$ gauge boson (right panel). 
The conventions for the various lines are the same as in Fig.~\protect{\ref{fig:susy}}. 
In the UED case (left), we indicate with a dotted black line the pair annihilation cross section of the LKP as a function of the mass, 
and shade in green the most conservative mass range where the LKP might give the WMAP inferred CDM abundance within 2-$\sigma$. 
In the right panel, the green shaded area corresponds to LZP realizations giving a relic abundance consistent with the upper limit on the CDM abundance.}
\label{fig:extradim}
\end{figure}

The flux of particles produced in the galactic halo from WIMP pair annihilations is always proportional to the product of the 
number density of WIMPs squared (giving the actual number of WIMP pairs) times the pair annihilation rate, $\sv$. 
A convenient representation of the sensitivity of indirect DM search experiments is therefore the plane $(m_{\chi},\sv/m_\chi^2)$. 
We assess the sensitivity of antideuteron searches in that plane, for various WIMP DM setups, in Fig.~\ref{fig:susy} and \ref{fig:extradim}. 

It goes without saying that the antideuteron fluxes are highly correlated with the antiproton fluxes, 
and, in view of the wealth of experimental results on the flux of antiprotons\cite{bess1,bess2,capricepbar}, this gives a possible constraint on the WIMP models under consideration. In the case of positrons, the correlation with the antideuteron flux is less straightforward, and in some cases does not hold.
However, it was pointed out in Ref.~\cite{Profumo:2004at} that the positron and antiproton fluxes indeed are typically correlated, at least at low energies. 
Given a WIMP setup, one can therefore require, as well, consistency with positron flux measurements\cite{mass91,heat,capriceeplus} (We adopt here, for the positron propagation, the approach outlined in Ref.~\cite{Baltz:1998xv}; see also the discussion in Sec.~3.1 of Ref.~\cite{Profumo:2004ty}).

Adopting a very conservative approach, one may ask that the primary antiproton and positron fluxes {\em alone} (i.e. neglecting the background: taking it into account would give stronger constraints) do not exceed the experimentally measured values. 
We consider here the antiproton flux data from BESS-98\cite{bess1,bess2} and CAPRICE-98\cite{capricepbar}, and the positron flux data from  MASS-91\cite{mass91}, 
HEAT-94/95\cite{heat} and CAPRICE-98\cite{capriceeplus}. 
Indicating the experimental data as $(E_i,\phi^{\rm exp}_i,\Delta\phi^{\rm exp}_i)$, 
and the primary antimatter flux after diffusion and solar modulation at an energy $E$ as $\phi^{\chi\chi}(E)$, we define the quantity
\begin{equation}\label{eq:xidef}
\xi={\rm Max}_i\left(\frac{\phi^{\chi\chi}(E_i)}{\phi^{\rm exp}_i+2\times\Delta\phi^{\rm exp}_i}\right).
\end{equation}
If $\xi>1$, the WIMP model is excluded at 95\% C.L., even assuming a negligible secondary and tertiary background.

A typically stronger constraint can be found by adding an independently calculated secondary antiproton or positron background on top of the 
primary supersymmetric contribution (for details on the computation of the secondary antiproton and positron background we use in the present analysis, see Ref.~\cite{Profumo:2004ty}). 
One can then require that the resulting $\chi^2$ to the measured data is statistically acceptable. 
Uncertainties in the computation of the standard antiproton background (see e.g.~\cite{Donato:2001sr}, 
where it is estimated a 20\% uncertainty from the diffusion model and 20\% from the nuclear cross sections) should be kept in mind when considering this constraint.

We show our results for WIMP models annihilating with branching fraction equal to 1 into $b\bar b$ pairs (left panel) and $W^+W^-$ pairs (right panel) in Fig.~\ref{fig:susy}.
We pick these two particular final states for a number of reasons: 
first, they respectively give the largest and smallest possible antideuteron yields among non-leptonic final states, 
so our results can be regarded as (model-independent) upper and lower limits for WIMP detection through $\dbar$ searches\footnote{If a WIMP annihiates into purely leptonic channels with a sizeable branching ratio, this effect must be properly taken into account when considering the right panel of Fig.~\ref{fig:susy} as a lower limmit.}. 
Secondly, these two final states constitute a benchmark for a number of neutralino DM configurations within the MSSM. 
In particular, the cases of wino-like neutralinos in the mAMSB scenario \cite{Giudice:1998xp,Randall:1998uk,Gherghetta:1999sw,Feng:1999hg}, 
and of higgsino-like DM \cite{hb_fp,Baer:2005ky} will closely resemble the $W^+W^-$ case, while many bino-like configurations, 
including the well-known bulk\cite{bulk} and $A$-funnel\cite{Afunnel} regions of minimal supergravity\cite{msugra}, 
will follow the sensitivity patterns outlined in the $b\bar b$ case.

The constraints from the absolute 95\% C.L. upper limit (i.e. in the limit of negligible background, see Eq.~\ref{eq:xidef}) on the positron and antiproton fluxes exclude the regions above the (upper and, respectively, lower) 
solid blue lines. Requiring an overall consistency of the primary component plus the computed background with all the available data amounts to 
ruling out the portions of the ($m_\chi,\sv/m_\chi^2$) plane lying above the dotted blue lines. 

Turning to antideuteron searches, the BESS upper limit on the $\dbar$ flux is indicated with a black line. 
In both cases, the region of parameter space excluded by the BESS results is already ruled out by current data on antiproton and positron data, 
and therefore no extra constraints can be derived.

The projected AMS sensitivity is indicated with red dashed lines: the upper, thicker, line indicates the pair annihilation cross section needed to produce a number of primary $\dbar$s which would entail, on average, the possibility of a discrimination of the signal over the background at the 95\% C.L., according to the criterion outlined in Sec.~\ref{sec:bckg}; the lower, fainter line indicates the threshold for the production of 1 $\dbar$ event at AMS of primary origin.

The sensitivity of the balloon-borne version of GAPS is presented with magenta double-dotted-dashed lines. The background for this experiment is very low, and the detection of even only one $\dbar$ would indicate a primary component to a high confidence level.The upper line corresponds to a LDB mission (to be launched from Antarctica), while the lower line to an ULDB mission from Australia. 

Finally, the dot-dashed green lines (again with the same convention for the fainter and thicker lines as for AMS) correspond to the projected sensitivity of GAPS on a satellite orbiting at high latitude around Earth\cite{Mori:2001dv}, 
and on an interplanetary deep space probe (in this case, we use the same sensitivity quoted in \cite{Mori:2001dv}, but with the interstellar, 
instead of solar modulated, $\dbar$ flux \cite{Profumo:2004ty}). 

To make contact with realistic MSSM setups, and as a guideline, we shade in yellow the absolute upper limit on the quantity $\sv/m_\chi^2$ 
derived in Ref.~\cite{Profumo:2005xd}, while the orange shaded area corresponds to supersymmetric configurations featuring a WMAP 
thermal neutralino relic abundance (see again Ref.~\cite{Profumo:2005xd} for details).

The consequences of the trend in the low-energy antideuteron flux for the gauge bosons final state outlined in Fig.~\ref{fig:dnde} is manifest in the right panel of Fig.~\ref{fig:susy}. 
Increasing the WIMP mass, the $W^+W^-$ yield in the low energy region tends to deplete, and, as a consequence, the annihilation rate needed to detect 
at least one $\dbar$ is increased: gauge bosons final states produce energetic antiprotons and antineutrons, hence less, and more energetic, antideuterons. Further, in the $W^+W^-$ final state case, the advantages of performing a deep space probe GAPS mission would be missed: 
the solar modulation effects, in fact, shifts the $\dbar$ spectrum towards lower energies, hence replenishing the low fluxes of low-energy antideuterons 
generated by the hadronic decays of the $W^+W^-$ final state. We wish to point out, however, that the predictions for the $\dbar$ flux at masses around 100 GeV are, 
in a sense ``universal'': the rates of $\dbar$ production for the most and least efficient channels, around that particular value of the WIMP mass, are, in fact, very close.

We also notice that the $\dbar$ production will be largely correlated to the neutralino annihilation induced nucleosinthesis of primordial $^6$Li, as estimated in Ref.~\cite{Jedamzik:2004ip}. In particular, it was shown in Ref.~\cite{Jedamzik:2004ip} that, for the particular quark-antiquark final state of Fig.~\ref{fig:susy}, left, neutralinos with $\sv/m_\chi^2\sim10^{-29}\div10^{-30}\ {\rm cm}^3{\rm s}^{-1}{\rm GeV}^{-2}$ would produce an amount of primordial $^6$Li consistent with observations \cite{li6obs}. This entails the intriguing consequence that if low energy antideuterons from neutralino annihilations are detected, {\em e.g.} by the ULD GAPS/balloon experiment (where a very low astrophysical background is expected, see Sec.~\ref{sec:bckg}), then a sizeable fraction of the observed $^6$Li might have been sinthesized in neutralino pair annihilations. The vice-versa, however, does not hold in general: for a dominant gauge bosons final state (Fig.~\ref{fig:susy}, right) one could produce the right amount of $^6$Li without implying a detectable flux of low energy antideuterons. Again, this depends upon the spectral distribution of $\dbar$s in this particular final state, see Fig.~\ref{fig:dnde}.

The same approach to the determination of the sensitivity of DM detection in future $\dbar$ search experiments, 
and of parameter space constraints from the BESS data and from antiproton and positron flux measurements, 
is applied in Fig.~\ref{fig:extradim} to two particular extra-dimensional setups featuring a DM candidate. 
In the left panel, we consider the minimal UED model\cite{Appelquist:2000nn,Servant:2002aq}, and, together with the above described sensitivity lines, 
we show the pair annihilation cross section over the mass squared, $\sv/m_{\rm LKP}^2$ for a $B^{(1)}$ LKP. 
The green shaded strip corresponds to the most conservative possible range of masses giving rise to a relic LKP abundance compatible with the WMAP 95\% C.L. 
range for the CDM abundance. 
In particular, recent evaluations of the relic abundance in UED scenarios include Ref.~\cite{Kong:2005hn,Burnell:2005hm}, 
where all coannihilation channels have been taken into account, and \cite{Kakizaki:2005en,Kakizaki:2005uy}, where resonant annihilations 
through $n=2$ KK excitations were considered. We point out that, quite remarkably, 
{\em the entire WMAP-allowed region will produce at least one primary antideuteron at a satellite-based GAPS-like experiment}. Since the estimated background for GAPS/satellite is of 1.7 events, in the present setup, we also point out that most of the WMAP compatible parameter space of the UED model will give a signal-to-background ratio larger than 1 at that future experiment.
On the other hand, provided some relic density enhancement mechanism (including a modified quintessential cosmology \cite{quint}, 
a Brans-Dicke-Jordan cosmology \cite{Kamionkowski:1990ni,bdj}, 
a primordial anisotropic expansion resulting in a non-vanishing shear energy density \cite{Kamionkowski:1990ni,Profumo:2004ex}, 
a brane-world scenario such as that proposed in Ref.~\cite{Nihei:2005qx}, or non-thermal particle production \cite{nonth}) are in place, 
lower LKP masses can also produce a sizable CDM density, and be detected at AMS or at balloon-borne GAPS missions.

In the right panel we explore the case of the LZP, assumed to be a stable Kaluza-Klein Dirac right-handed neutrino in the context of a 5-dimensional 
warped extra dimensional GUT scenario\cite{Agashe:2004ci,Agashe:2004bm,Hooper:2005fj}. 
As pointed out in \cite{Agashe:2004bm,Hooper:2005fj}, there are mainly two regions allowed by the requirement of producing a WMAP relic abundance. 
The first one is for LZP masses close to the $Z$ pole, where the LZPs mainly annihilate through the resonant $s$-channel $Z$ boson, 
while the second one appears at larger masses (in a range which is largely model-dependent) where the $t$ channel exchange of GUT gauge bosons 
efficiently reduces the LZP relic abundance. 
Since the case of a WIMP dominantly pair-annihilating through the $Z$ can be considered a benchmark in other scenarios as well\cite{Jungman:1995df}, 
we decide to focus here on the low-mass range of viable LZP masses. 
We shade in green, in Fig.~\ref{fig:extradim} (right panel), the range on the $(m_\chi,\sv/m_\chi^2)$ plane where a sufficiently large 
$\sv\gtrsim 10^{-26}\ {\rm cm}^3/{\rm s}$ is achieved. 
In the lower-central area of the shaded region, the relic abundance will be somewhat lower than the 95\% C.L. limit: 
in this case either one invokes, again, a relic abundance enhancement mechanism, or, alternatively, one has to rescale the values of $\sv$ 
according to some procedure, for instance multiplying it by a factor $\epsilon^2$, where $\epsilon={\rm min}[1,(\Omega_\chi h^2/0.09)^2]$. 
In the first scenario (no rescaling, relic density enhancement) the full range of low-mass LZPs will be within even the LDB GAPS mission; 
had we rescaled $\sv$ as described above, a narrow mass range would escape the detection at future $\dbar$ searches, 
but most of the parameter space would still be accessible to a satellite based GAPS mission.

As a last comment, we point out that the AMS sensitivity, taking into account the issue of the background, is in most cases worse than even an LDB GAPS experiment, and it is often incompatible with the constraints from $\overline{p}$ fluxes. These conclusions naturally jeopardize the scientific opportunities of dark matter searches through low-energy antideuterons with AMS, and strenghten and further motivate the need for a GAPS-like experiment to pursue this dark matter search technique.

\subsection{Comparing $\dbar$ detection to other direct/indirect WIMP searches}
\label{sec:compare}

\begin{figure}[!t]
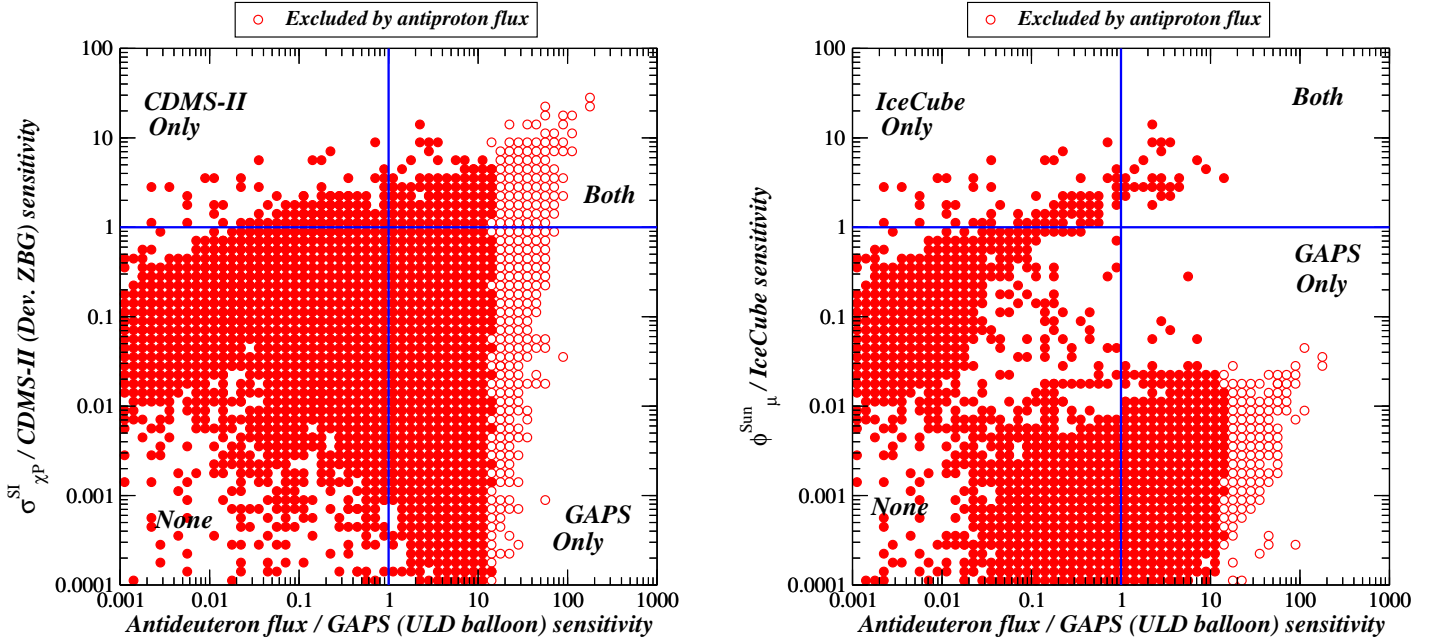

\begin{center}
\hspace*{-0.7cm}\mbox{\epsfig{file=plots/bal_cdms.eps,height=8.5cm}\quad\quad\epsfig{file=plots/bal_ice.eps,height=8.5cm}}
\end{center}
\caption{\it\small 
({\em Left panel}): The correlation between the expected sensitivity of antideuteron searches and of direct detection experiments for 
supersymmetric models giving a thermal neutralino relic abundance in the WMAP range. On the $y$ axis we indicate the ratio of the 
neutralino-proton spin-independent scattering cross section over the projected maximal sensitivity of the CDMS-II experiment, 
at the WIMP mass corresponding to the neutralino mass for the model under consideration. 
On the $x$ axis we indicate the number of expected primary antideuterons detected at the ULDB GAPS mission 
(ratio of the average flux over the experimental sensitivity). 
({\em Right panel}): the same as in the left panel, but correlating the sensitivity of antideuteron searches at GAPS on a ULDB mission with that 
IceCube for the neutralino-annihilation induced flux of muons from the center of the Sun.}
\label{fig:balloon}
\end{figure}

An important issue in evaluating the role of antideuteron searches in the quest for a DM signal is to compare the reach of this 
detection channel with other detection techniques. We carry out this task in Fig.~\ref{fig:balloon}, 
where we compare the sensitivity of the ultimate CDMS-II apparatus\cite{Brink:2005ej} with the expected sensitivity of GAPS on an ULDB mission\cite{hailey_privcom}. 
We analyze the particular case of neutralino dark matter, and perform a scan of the general MSSM along the lines of Ref.~\cite{Profumo:2004at} (The details of the scan are provided in Tab.~\ref{tab:scan}). 
We only consider neutralino DM models giving a thermal relic abundance within the WMAP range, $0.09\lesssim\Omega_\chi h^2\lesssim0.13$ \cite{Spergel:2003cb}, so we do not include here low relic density models. 
We plot on the $x$-axis the number of expected primary antideuterons to be detected by GAPS, and on the $y$-axis the ratio, 
at the particular neutralino mass of the model under consideration, the neutralino-proton spin-independent scattering cross section 
over the experimental projected sensitivity. Models lying above the horizontal blue line will be detectable at CDMS-II, 
while models to the right of the vertical blue band lie within the ULDB GAPS/balloon sensitivity. 
Models giving rise to excessive antiproton fluxes ({\em i.e.} $\xi>1$ according to the definition given in Eq.~(\ref{eq:xidef})) are indicated with empty circles. 
If a model is indicated in the upper right area, it will be detectable at both GAPS and CDMS-II.

\begin{table}[!b]
\begin{center}
\begin{tabular}{|c|c|c|c|c|c|c|c|}\hline
$\mu$ & $m_1$ & $m_2$ & $m_3$ & $m_A$ & $m_{\widetilde S}$ & $A_{\widetilde S_3}$ & $\tan\beta$\\
\hline
$50\div2000$ &  $2\div2000$ &  $80\div2000$ &  $m_{\rm\sss LSP}\div20000$ &  $100\div10m_{\rm\sss LSP}$ &  $(1\div10)m_{\rm\sss LSP}$ &  $(-3\div3)m_{\widetilde S}$ &  $1\div60$\\
\hline
\end{tabular}
\end{center}
\caption{\it\small Ranges of the MSSM parameters used to generate the models shown in Fig.~\protect{\ref{fig:balloon}} and Fig.~\protect{\ref{fig:satellite}}. All masses are in GeV, and $m_{\rm\sss LSP}\equiv{\rm min}(\mu,m_1,m_2)$. $m_{\widetilde S}$ indicates the following scalar masses (which were independently sampled): $m_{\widetilde Q_{1,3}}$, $m_{\widetilde u_{1,3}}$, $m_{\widetilde d_{1,3}}$, $m_{\widetilde L_{1,2,3}}$, $m_{\widetilde e_{1,2,3}}$. To avoid FCNC constraints, we assumed the squark soft supersymmetry breaking terms of the first two generations to be equal. $A_{\widetilde S_3}$ stands for the third generation sfermions trilinear terms: those of the first two generations were taken to vanish.}\label{tab:scan}
\end{table}
We wish to point out the complementarity of the two detection techniques: many models which will not give a large enough direct detection 
signal will instead be ``detectable'' at GAPS. The vice versa also holds, although we do not find many models that can be visible at CDMS-II 
while not giving a significant antideuteron flux. We also notice that, within supersymmetric models, 
the maximal number of primary antideuterons one can expect to detect with a ULDB GAPS mission is between 10 to 20 antideuterons: 
a larger number of $\dbar$s is excluded by current antiproton constraints.

In the right panel, we compare the sensitivity of GAPS/balloon with that of the ${\rm km}^3$-sized detector IceCube, 
looking for muons produced by neutrinos originating from neutralino pair annihilations in the core of the Sun\cite{icecube,icecube_sens}. 
Again, the two detection techniques are somewhat complementary, and a few supersymmetric models might give a signal at both facilities. Notice that the points approximately cluster in two distinct regions, which correspond to models where the capture/annihilation equilibrium inside the Sun is, or not, achieved.

\begin{figure}[!t]
\begin{center}
\epsfig{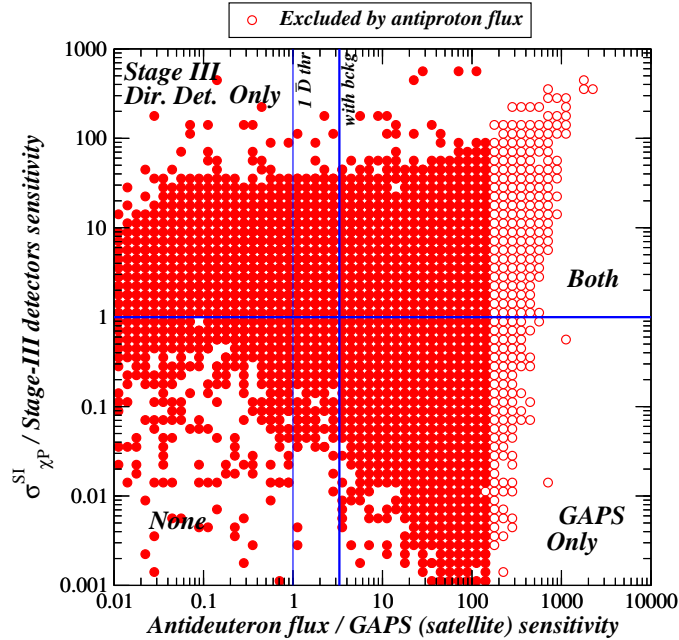}
\end{center}
\caption{\it\small 
Same as in Fig.~\protect{\ref{fig:balloon}}, but correlating the sensitivity of Stage-III direct detection experiments 
(in particular, the Xenon 1-t facility) with the sensitivity of GAPS on a satellite-borne mission. The fainter vertical blue line to the left indicates the detection threshold for 1 primary $\dbar$, while the thicker line to the right indicates the threshold for 3.3 primary $\dbar$s, corresponding to a statistical detection of a primary component over the background at the 95\% C.L. (see the discussion in Sec.~\ref{sec:bckg}).}
\label{fig:satellite}
\end{figure}
In Fig.~\ref{fig:satellite}, finally, we compare the detection capabilities of ton-sized direct detection experiments 
(employing as a benchmark experimental setup the case of Xenon-1t \cite{xenon}) with the prospects of antideuteron detection of a 
GAPS mission on a satellite \cite{Mori:2001dv}. Most of the supersymmetric models included in our scan will be detected at one of the two experiments, 
and a sizable fraction will give a signal at both experiments. The increased sensitivity of GAPS will allow one to collect a number of 
primary antideuterons as large as 100-200, without conflict with antiproton flux measurements.

\subsection{Uncertainties in the antideuteron flux computation}\label{sec:unc}

\begin{figure}[!t]
\begin{center}
\epsfig{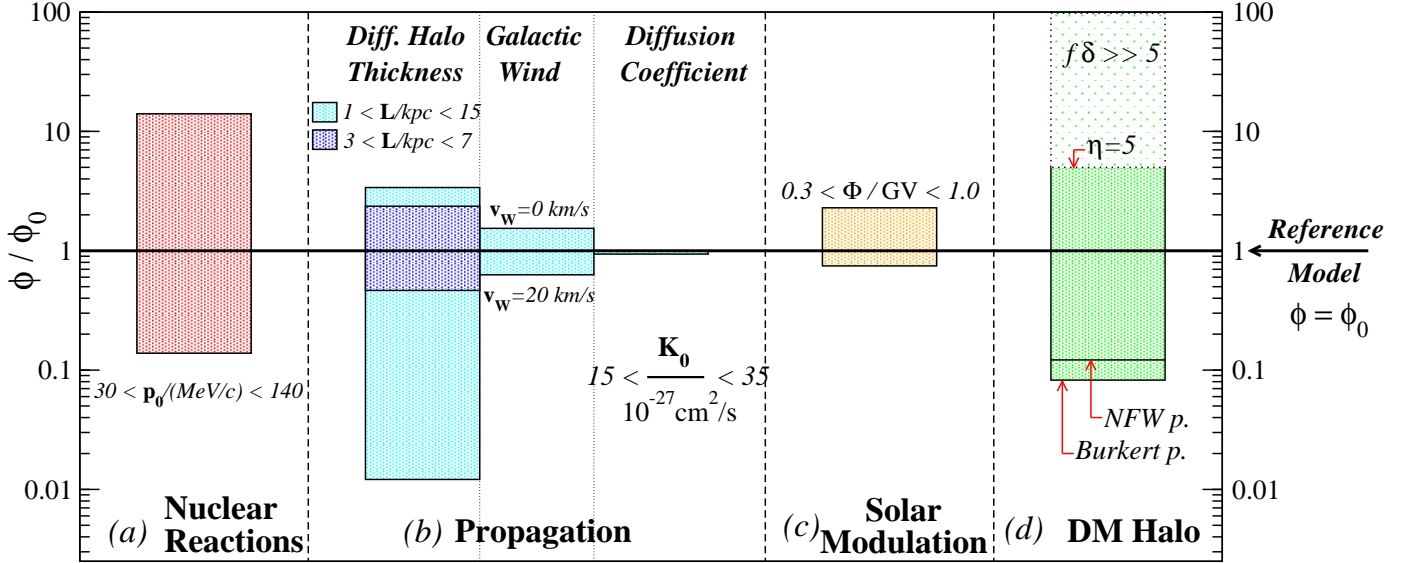}
\end{center}
\caption{\it\small 
A compilation of the various sources of uncertainty in the primary antideuteron flux computation, 
expressed as the ratio of the antideuteron flux in the models under consideration over that of a ``reference setup'', defined in the text. 
Panel (a) shows the nuclear reaction uncertainty parameterized by the variation of the coalescence momentum $p_0$ within the range 
$30<p_0/{\rm MeV/c}<140$. Panel (b) highlights the uncertainty due to the diffusion of the antideuterons in the galactic halo, 
induced by the diffusion halo thickness $L$, which was varied within the range $1<L/{\rm kpc}<15$ and within the (more ``realistic'') 
range $3<L/{\rm kpc}<7$ \protect{\cite{Moskalenko:2001ya}} (blue shaded column), 
by the galactic wind velocity $v_W$, varied in the range $0<v_W/{\rm km/s}<20$, and by the diffusion coefficient $K_0$, 
varied in the range $15<K_0/(10^{-27} {\rm cm}^2{\rm /s})<35$. 
Panel (c) shows the variation of the flux with the solar activity, due to modulation of the antideuteron flux in the heliosphere. 
The modulation parameter $\Phi$ was varied between 300 and 1000 MV. 
Finally, panel (d) sketches the uncertainty originating from the structure of the DM halo, 
showing the resulting flux variation for two well-known halo profiles, the NFW profile\cite{nfw} and the Burkert profile\cite{salucci-burkert}, 
and the enhancement factor $\eta$ due to DM clumps, as computed in Ref.~\cite{Berezinsky:2003vn}. 
The light shaded area above the $\eta=5$ line reminds the possibility of a larger boost factor from halo clumpiness\cite{Bergstrom:1998jj}.}
\label{fig:unc}
\end{figure}

The computation of the cosmic ray yields resulting from DM pair annihilations in the galactic halo is plagued by a number of uncertainties, 
which we wish to discuss and compare, when possible, here. 
To this extent, we consider a WIMP model pair annihilating with BR=1 into a $b\bar b$ pair, a mass $m_\chi=100$ GeV and a pair 
annihilation cross section $\sv=3\times 10^{-26}\ {\rm cm}^3/{\rm s}$. 
The nuclear and propagation model parameters we use are the same as those discussed in the preceding section. 
We stress that the details of the particle physics model are not critical in the determination of the uncertainties we want to assess here. 
We show in Fig.~\ref{fig:unc} the ``uncertainty factor'', defined as the ratio of the flux computed varying the setup with respect to the reference one, 
over the reference antideuteron flux. 
We compute, for definiteness, the $\dbar$ flux at a kinetic energy per nucleon of 0.2 GeV, relevant for all the $\dbar$ search experiments under consideration here.

First, in the case of nuclei production, a major source of ambiguity lies in the nuclear reaction cross sections involved in the particular process under consideration. 
For antideuterons, the antiproton-antineutron fusion is described, as outlined above, by the coalescence model, 
where the nuclear cross section uncertainty can be parameterized in terms of the coalescence momentum $p_0$. 
To quantify the resulting uncertainty band, we vary $p_0$ in the range quoted in Ref.~\cite{Barrau:2002mc}, $30\lesssim p_0/{\rm (MeV/c)}\lesssim 140$ 
with our definition of $p_0$, which is consistent with the available data on the antideuteron production cross section from accelerator experiments\cite{Albrecht:1989ag}. 
We find that the total uncertainty spans around two orders of magnitude, the reference value used in the computations above 
giving a reasonable central average of the possible outcomes.

A second major source of uncertainty stems from the propagation of the charged cosmic ray species under consideration here through the Galactic magnetic fields. 
As pointed out in \cite{Barrau:2002mc}, the largest effects come from a variation of the half-height $h_h$ of the halo diffusive region. 
We vary the parameter $h_h$ both in the wide range suggested in \cite{Barrau:2002mc}, $1\lesssim h_h/{\rm kpc} \lesssim 15$ 
(light blue band; notice that the diffusion model used in that paper differs from the one we consider here, so this range might be an over-estimate of the physically allowed one) 
and in the range suggested in \cite{Moskalenko:2001ya}, $3\lesssim h_h/{\rm kpc} \lesssim 7$, 
which applies to a propagation model qualitatively very similar to the one we use here (blue shaded band). 
The induced spread in the $\dbar$ flux is remarkably large, but restricting to the range quoted in Ref.~\cite{Moskalenko:2001ya} 
the overall relative uncertainty factor is around a factor 5. The largest fluxes are obtained at larger values of the diffusive zone, and vice-versa. 

A variation of the galactic wind velocity, in the range $0\lesssim v_W/{\rm (km/s)}\lesssim20$, induces an uncertainty factor around 2. 
The larger the galactic wind velocity, the larger the number of antideuterons leaking outside the diffusive region, and, 
therefore, the smaller the final top of the atmosphere $\dbar$ flux. 
Finally, we find that a variation of the diffusion coefficient $K_0$ affects very mildly the low-energy antideuteron flux.

Turning to the role of the solar modulation effects, we include here an assessment of the variations induced by a change in the Fisk parameter $\phi_F$, 
although the latter can, in principle, be estimated from other cosmic ray species fluxes\cite{Bottino:1998tw}, for a given period of time, in which case the induced uncertainty would largely be under control. 
As $\phi_F$ is varied from 0.3 to 1 GV, the low-energy $\dbar$ flux change within a factor around 3.

All the above considered sources of uncertainty affect both the computation of the primary $\dbar$ flux and that of the $\dbar$ ``background''. 
As recently pointed out in Ref.~\cite{Duperray:2005si}, the main sources of uncertainty in the astrophysical background computation are, 
in decreasing order of magnitude, the hadronic cross sections, solar modulation and propagation. 
This is consistent with the present analysis, although we find that $\dbar$ propagation is likely giving a larger effect than solar modulation.

Finally, the primary $\dbar$ flux (but {\em not} the secondary and tertiary background) depends on the assumed structure of the Dark Matter halo. 
In the case of a smooth halo, a critical quantity entering the computation of the primary cosmic ray flux is the local DM halo density. The latter cannot be taken as a free parameter, but has to be chosen consistently with the halo profile and the related observational constraints. 
In addition, it is a subject of debate as to how the actual halo shape, in its innermost regions, 
affects the abundance of primary particles produced in WIMP annihilations. 
It was pointed out in Ref.~\cite{Bergstrom:1999qv} that in the case of a cuspy NFW profile, up to 43\% (depending on the core radius) 
of the $\bar p$ arriving at Earth are produced in a sphere of radius 1 kpc around the galactic center, 
while only 1\% in the case of a shallower halo, as for example the isothermal sphere profile. 
We find quite significant differences by resorting to a NFW \cite{nfw} or a Burkert profile \cite{salucci-burkert} (giving the lowest fluxes for the set of halo profiles we consider here), compared to the adiabatically contracted N03 halo model (our reference model). This is due both to the shape of the dark matter halo and to the local dark matter density, which for the three profiles at hand respectively reads, in units of GeV/${\rm cm}^3$, 0.3 (NFW), 0.34 (Burkert) and 0.38 (adiabatically contracted N03). Requiring self-consistency in the computation of the velocity profiles for the dark matter halos does not allow much freedom in the choice of the local halo density, but since the $\dbar$ flux approximately scales quadratically with $\rho_{\rm DM}$ one should keep in mind that the range of possible outcomes considering other local halo density figures might be marginally larger than what we quote here.

A second source of possible enhancement of the $\dbar$ flux comes from the possibility of the existence of clumps in the dark halo, 
which would create high DM density concentrations giving rise to a possibly significant increase in the number of WIMP annihilations. 
A model-independent approach to the effects of clumpiness in indirect DM detection is given in \cite{Bergstrom:1998jj}, 
where the relevant quantity driving the primary cosmic ray flux enhancement was shown to be $f\cdot\delta$, 
where $f$ is the fraction of dark matter forming clumps, while $\delta$ is a typical clump overdensity (for quantitative definitions see \cite{Bergstrom:1998jj}). 
In \cite{Bergstrom:1998jj}, it was pointed out that the possible clumpiness enhancement factors can in principle be as large as $10^9$, 
in the context of supersymmetric dark matter, without violating the bounds coming from antiproton and gamma-ray fluxes. 
More detailed and model-dependent results were recently given in Ref.~\cite{Berezinsky:2003vn}, where it was noticed that only a small fraction 
of small-scale clumps (less than 1\%) is likely to survive tidal destruction. 
It was also claimed that clumps are not cuspy, again due to tidal interactions (hence lowering the maximal overdensity $\delta$), and, 
resorting to a primeval fluctuation index close to 1 as preferred by observation, the enhancement factor $\eta$ was computed to be between 2 and 5. 
The computation relies however on many assumptions, and it can presumably be considered as a conservative scenario. 
To summarize, the overall uncertainty in the primary flux coming purely from the smooth component of the halo is around one order of magnitude, 
while that from the occurrence of clumpiness may in principle be very large, although the indicative range is within a factor 2$\div$5\cite{Berezinsky:2003vn}.

\section{Conclusions}\label{sec:conclusions}

We summarize below the main results of the present analysis:

\begin{itemize}
\item We showed that a recent re-evaluation of the secondary and tertiary antideuteron background jeopardizes the possibility of extracting a 
clean clue for new physics from $\dbar$ searches at AMS-02. The background, however, should be negligible for a balloon-borne GAPS mission, 
while a few events might be expected at a GAPS satellite mission.
\item We carried out a model independent analysis of the primary antideuteron flux expected from WIMP pair annihilation in the galactic halo, 
and compared it with the current and future sensitivity of $\dbar$ search experiments. 
The recently reported BESS upper limit on the $\dbar$ flux does constrain a few extreme WIMP setups, which, however, 
are already ruled out by current antiproton and positron data. 
Future balloon- and satellite-borne experiments looking for low-energy antideuterons produced in DM pair annihilations will be able to access 
large portions of the supersymmetric parameter space, and will be sensitive to signals from various DM models in extra-dimensional scenarios. 
In particular, the GAPS experiment on a satellite is found to be sensitive to primary antideuterons, with a signal to background typically larger than 1, over the whole WMAP compatible parameter space of the minimal UED model.
\item We outlined a significant complementarity between antideuteron searches and other DM search techniques, in particular direct detection. 
We pointed out that consistency with currently available antiproton flux measurements implies an upper bound on the maximal number of 
primary antideuterons which can be detected at GAPS on a ULDB mission (GAPS on a satellite) of around 20 (200) events.
\item We highlighted, and quantified, the various sources of uncertainty in the primary antideuteron flux computation, 
ranging from the parameterization of nuclear processes, to the propagation of antideuterons throughout the Galaxy, 
solar modulation effects and the structure of the dark matter halo.
\end{itemize}


\noindent{ {\bf Acknowledgments} } \\
\noindent
We gratefully acknowledge many interesting suggestions, comments and discussions with Chuck Hailey and Fiorenza Donato. We thank Alberto Belloni for various advices related to a few statistics-related issues, and Karsten Jedamzik and Piero Ullio for comments and discussions. This work was supported in part by the U.S. Department of Energy under contract number DE-FG02-97ER41022.


\end{document}